\documentclass[seceq]{ptptex}
\notypesetlogo                       %comment in if to eliminate PTPTeX 
\title{%        %You can use \\ for explicit line-break
A Brane World in an Arbitrary Number of Dimensions without $Z_{2}$ Symmetry%
}

\author{%       %Use \scshape  for the family name
Daisuke \textsc{Yamauchi}\footnote{E-mail: yamauchi@yukawa.kyoto-u.ac.jp} 
and Misao \textsc{Sasaki}\footnote{E-mail: misao@yukawa.kyoto-u.ac.jp} 
}

\inst{%         %Affiliation, neglected when [addenda] or [errata]
Yukawa Institute for Theoretical Physics, Kyoto University,\\ Kyoto 606-8502, Japan
}

\abst{%         %this abstract is neglected when [addenda] or [errata]
We consider a brane world in an arbitrary number of dimensions without $Z_2$ symmetry
and derive the effective Einstein equation on the brane, where its right-hand
side is given by the matter on the brane and the curvature in the bulk.
This is achieved by first deriving the junction conditions
for a non-$Z_2$ symmetric brane
and second solving the Gauss equation, which relates the mean
extrinsic curvature of the brane to the curvature in the bulk, with respect to
the mean extrinsic curvature. The latter corresponds
to formulating an explicit junction condition
on the mean of the extrinsic curvature, analogue to the Israel junction condition
for the jump of the extrinsic curvature.
We find that there appears a new type of 
effective anisotropic fluid on the right-hand side of
the effective Einstein equation due to the fact that there is no $Z_2$ symmetry.
The derived equation is a basic equation for the study of
Kaluza-Klein brane worlds in which some dimensions on the brane are compactified
or for a regularization scheme for a higher codimension brane world, where 
the Kaluza-Klein compactification on the brane is regarded as a means to
regularize the uncontrollable spacetime singularity created by the higher
codimension brane.
}

\begin{document}

\maketitle

\section{Introduction}
String theory suggests that our universe is not four dimensional 
but, rather, a submanifold (brane) embeded in a higher-dimensional
spacetime (bulk).
In particular, Radall and Sumdrum (RS)~\cite{Randall:1999vf,Randall:1999ee}
 proposed an interesting brane world model inspired by the Horava-Witten 
model \cite{Horava:1996ma}.
The RS model assumes a codimension-1 brane with $Z_{2}$ symmetry
embeded in the bulk with a negative cosmological constant.
However, to reconcile a higher-dimensional theory with the
observed four-dimensional spacetime, the RS model is not sufficient.
Since string theory suggests that the number of bulk dimensions
is 10 or 11, the corresponding number of codimensions is 6 or 7.
Therefore, we must consider a higher-codimension brane world.
But a higher-codimension brane world has the serious problem that the brane
becomes an uncontrollable spacetime singularity due to its self-gravity,
except possibly for a codimension-2 brane world, which may give a reasonable
cosmology~\cite{Cline:2003ak}.
Thus it is necessary to develop a regularization method to realize 
a reasonable higher-codimension brane world.

For the above-stated purpose, we focus on a specific regularization scheme,
which we now describe.
Let us consider a codimenion-$(q+1)$ brane in an $n$-dimensional
spacetime. We regularize this brane by expanding it into $q$-dimensions,
so that it becomes a codimension-1 brane with
$q$ compact dimensions on the brane. In 3 dimensions, 
for example, this correponds to replacing a point by a small 2-sphere.
Note that the resulting codimension-1 brane will not have the $Z_2$ symmetry.

This regularization scheme is essentially the same as
the Kaluza-Klein (KK) compactification of $q$ spatial dimensions on
the brane, which is called the KK brane world~\cite{Chatillon:2006vw}.
In the codimension-2 case, Peloso et al.~\cite{Peloso:2006cq} considered 
KK brane worlds with the internal two-dimensional space compactified
by a flux (see also Refs.~\citen{Kobayashi:2007kv,Papantonopoulos:2006dv,Himmetoglu:2006nw}). 
They showed using a perturbative analysis that four-dimensional
Einstein gravity is recovered on the brane.
Recently, Kanno et al.~\cite{Kanno:2007wj} investigaged
the cosmology of the KK brane worlds in a non-trivial bulk background.

In this paper, partly to give a framework for the 
KK brane worlds and partly as a first step to formulate
the above-mentioned regularization scheme for
brane worlds of arbitrary codimension,
 we consider a codimension-1 brane world in an arbitrary number of
spacetime dimensions without $Z_2$ symmetry and derive an effective
Einstein equation on the brane, which is a generalization of the
effective Einstein equation on the brane with $Z_2$ symmetry derived
by Shiromizu, Maeda and Sasaki~\cite{Shiromizu:1999wj}.

There has been some work on non-$Z_2$ symmetric 
branes.~\cite{Battye:2001yn,Gergely:2003pn,Maia:2004fq,Ida:1999ui,Kanno:2005vq}
 The work most relevant to the present one is that of Battye
et al.,~\cite{Battye:2001yn} in which the non-$Z_2$ symmetric brane world
is investigated. They study the junction condition
in detail and point out that the effective Einstein equation 
has terms involving the mean of the extrinsic curvature across
the brane which are not explicitly expressed in terms of either
the matter on the brane or the curvature in the bulk.
Then, they focused their investigation on a spatially homogeneous,
isotropic brane. Our purpose here is to solve this problem and
express the effective Einstein equation solely in terms of the matter
on the brane and the curvature in the bulk, and also to present a straightforward
generalization in which the number of spacetime dimensions of
the bulk is extended from 5 to $n$.

This paper is organized as follows. In \S~\ref{sec:II},
we review the basic equations, in particular we give the junction condition
for non-$Z_2$ symmetric branes in Eq.~(\ref{eq:2ndGauss eq.}), first obtained by
Battye et al.~\cite{Battye:2001yn} in five dimensions and generalized
to arbitrary dimensions by Chatillon et al.~\cite{Chatillon:2006vw}.
We present the effective Einstein equation, but with terms which
still involve the mean of the extrinsic curvature,
which we call the pre-effective Einstein equation.
In \S~\ref{sec:III}, using the Gauss equation,
we solve for the mean of the extrinsic curvature across
the brane in terms of the matter on the brane and the curvature
in the bulk. \S~\ref{sec:IV} is devoted to a
conclusion and discussion.

Thoughout the paper, we use square brackets to denote the jump 
of a quantity across the brane and angled brackets to denote its mean.
For an arbitrary tensor ${\cal A}$ (with tensor indices suppressed),
we define 
\begin{eqnarray}
&&[{\cal A}]\equiv {\cal A}^{+}-{\cal A}^{-}\,,
\label{eq:[] def}
\\
&&\langle {\cal A}\rangle \equiv \frac{1}{2}({\cal A}^{+}+{\cal A}^{-})\,,
\end{eqnarray}
where the superscript $+$ denotes the side of the brane from which 
the normal vector $n^A$ (to be introduced in \S~\ref{sec:II})
points toward the bulk. We note that the brackets 
$\langle{\cal A}{\cal B}\rangle$ and $[{\cal A}{\cal B}]$
for the products of two tensors ${\cal A}$ and $\cal B$
can be expressed as
\begin{eqnarray}
&&\langle {\cal A}{\cal B}\rangle 
=\langle {\cal A}\rangle \langle {\cal B}\rangle 
+\frac{1}{4}[{\cal A}][{\cal B}]\,,
\label{eq:<>eq}
\\
&&\ [{\cal A}{\cal B}]=\langle {\cal A}\rangle [{\cal B}]
+[{\cal A}]\langle {\cal B}\rangle\,.
\label{eq:[]eq}
\end{eqnarray}

\section{Junction condition for $[K_{ab}]$}
\label{sec:II}

\subsection{$(n-1)+1$ decomposition}
We consider a family of $(n-1)$-dimensional timelike hypersurfaces
(slicing) in an $n$-dimensional spacetime and identify
one of them as a brane (i.e., a singular hypersurface).
We denote the bulk metric by $g_{MN}$ with the
coordinates $x^M$, where $M=0,1,\cdots,n-1$.
We denote the vector field unit normal to the hypersurfaces by
$n^M$. Then the induced metric $\gamma_{MN}$ on the hypersurfaces
is given by
\begin{eqnarray}
\gamma_{MN}=g_{MN}-n_Mn_N\,.
\end{eqnarray}

We express the internal coordinates on the brane by $y^{a}$, with
$a=0,1,\cdots,n-2$. The brane is embeded in the bulk as $x^{M}=x^{M}(y^{a})$.
Then, the induced metric on the brane
is expressed as
\begin{eqnarray}
&&\gamma_{ab}
=g_{MN}\frac{\partial x^{M}}{\partial y^{a}}
 \frac{\partial x^{N}}{\partial y^{b}}\,,
\nonumber\\
&&\gamma^{AB}
=\gamma^{ab}\frac{\partial x^{M}}{\partial y^{a}}
\frac{\partial x^{N}}{\partial y^{b}}\,.
\end{eqnarray}
The metric $\gamma^{A}{}_{B}=\gamma^{AC}g_{CB}$
acts as a projection operator, projecting bulk tensors onto the brane.

The extrinsic curvature $K_{ab}$ of the brane is defined as
\begin{eqnarray}
K_{ab}\equiv\frac{1}{2}\pounds_{n}\gamma_{ab}
=\gamma^{M}_{b}\gamma^{N}_{a}\nabla_{N}n_{M}\,.
\end{eqnarray}
Denoting the $n$-dimensinal Riemann tensor by ${\cal R}_{ABCD}$
and the $(n-1)$-dimensional Riemann tensor by $\bar{R}_{abcd}$,
the Gauss-Codazzi equations are expressed as
\begin{eqnarray}
&&{\cal R}_{ABCD}\gamma_{a}^{A}\gamma_{b}^{B}\gamma_{c}^{C}\gamma_{d}^{D}
=\bar{R}_{abcd}+2K_{a[d}K_{c]b}\,,
\nonumber\\
&&{\cal R}_{MABC}n^{M}\gamma_{a}^{A}\gamma_{b}^{B}\gamma_{c}^{C}
=2\bar{\nabla}_{[c}K_{b]a}\,,
\label{eq:Codazzi eq.}
\end{eqnarray}
where $\bar{\nabla}_{a}\equiv\gamma_{a}^{M}\nabla_{M}$ denotes
the covariant derivative associated with the brane induced metric 
$\gamma_{ab}$.
The bulk Riemann tensor is expressed in terms of the bulk Ricci tensor
and Weyl curvature ${\cal C}_{ABCD}$ as
\begin{eqnarray}
&&{\cal R}_{AB}{}^{CD}={\cal C}_{AB}{}^{CD}
+\frac{4}{n-2}g_{[A}{}^{[C}{\cal R}_{B]}{}^{D]}
-\frac{2}{(n-2)(n-1)}g_{A}{}^{[C}g_{B}{}^{D]}{\cal R}\,,
\end{eqnarray}
where ${\cal R}_{AB}\equiv R^{C}{}_{ACB}$ and
${\cal R}={\cal R}^{A}{}_{A}$.
Using this, the Gauss equation gives
\begin{eqnarray}
\bar{R}_{ab}&=&\frac{n-3}{n-2}{\cal R}_{AB}\gamma^{A}_{a}\gamma^{B}_{b}
+\frac{1}{n-2}{\cal R}_{CD}\gamma^{CD}\gamma_{ab}
-\frac{1}{n-1}{\cal R}\gamma_{ab}\nonumber\\
&&\ \ \ \ \ \ \ \ \ \ \ \ \ \ \ \ \ \ \ \ \ \ \ \ +KK_{ab}-K_{a}{}^{c}K_{bc}+{\cal E}_{ab}\,,
\label{eq:1stGauss eq.}
\end{eqnarray}
where $\bar{R}_{ab}=\bar{R}^{c}{}_{acb}$, and ${\cal E}_{ab}$ is 
the projected Weyl curvature on the brane, defined by
\begin{eqnarray}
{\cal E}_{ab}\equiv {\cal C}_{ACBD}n^{C}n^{D}\gamma^{A}_{a}\gamma^{B}_{b}\,.
\end{eqnarray}

For convenience, we introduce the tensor ${\cal F}_{ab}$
defined by
\begin{eqnarray}
{\cal F}_{ab}\equiv \frac{n-3}{n-2}{\cal R}_{AB}\gamma^{A}_{a}\gamma^{B}_{b}
+\frac{1}{n-2}{\cal R}_{CD}\gamma^{CD}\gamma_{ab}
-\frac{1}{n-1}{\cal R}\gamma_{ab}+{\cal E}_{ab}\,.
\label{eq:F def}
\end{eqnarray}
We denote its trace and traceless parts by ${\cal F}$ and $\omega_{ab}$,
\begin{eqnarray}
{\cal F}&\equiv&{\cal F}^{a}{}_{a}=2{\cal R}_{AB}\gamma^{AB}-{\cal R}
=2{\cal G}_{MN}n^{M}n^{N}\,,
\label{eq:F trace}
\\
\omega_{ab}&\equiv&\frac{n-3}{n-2}({\cal R}_{AB}\gamma^{A}_{a}\gamma^{B}_{b}
-\frac{1}{n-1}{\cal R}_{AB}\gamma^{AB}\gamma_{ab})+{\cal E}_{ab}\,,
\label{eq:F traceless}
\end{eqnarray}
where ${\cal G}_{AB}={\cal R}_{AB}-\frac{1}{2}{\cal R}g_{AB}$ 
is the $n$-dimensional Einstein tensor.
Thus, ${\cal F}_{ab}$ is decomposed as
\begin{eqnarray}
{\cal F}_{ab}=\frac{{\cal F}}{n-1}\gamma_{ab}+\omega_{ab}\,.
\label{eq F_ab decomposition}
\end{eqnarray}

Using ${\cal F}_{ab}$, we can rewrite the Gauss equation~(\ref{eq:1stGauss eq.})
as
\begin{eqnarray}
\bar{R}_{ab}={\cal F}_{ab}+KK_{ab}-K^{c}{}_{a}K_{bc}\,.
\end{eqnarray}
Using the fact that the brane induced metric satisfies 
the junction condition $[\gamma_{ab}]=0$,
the Gauss equation can be decomposed into two equations,
\begin{eqnarray}
\langle\bar{R}_{ab}\rangle &&=\bar{R}_{ab}
=\langle {\cal F}_{ab}\rangle +\langle KK_{ab}\rangle
 -\langle K^{c}{}_{a}K_{bc}\rangle
\nonumber\\ 
&&=\langle {\cal F}_{ab}\rangle +\frac{1}{4}([K][K_{ab}]-[K_{a}^{c}][K_{bc}])
+\langle K\rangle\langle K_{ab}\rangle
 -\langle K_{a}^{c}\rangle\langle K_{bc}\rangle\,,
\label{eq:<Gauss eq.>}\\
\ [\bar{R}_{ab}]&&=0=[{\cal F}_{ab}]+[KK_{ab}]-[K^{c}{}_{a}K_{bc}]
\nonumber\\
&&\qquad
=[{\cal F}_{ab}]+\langle K\rangle [K_{ab}]
+[K]\langle K_{ab}\rangle -2\langle K_{(a}{}^{c}\rangle [K_{b)c}]\,,
\label{eq:Gauss eq.}
\end{eqnarray}
where we have used the two product relations (\ref{eq:<>eq})
and (\ref{eq:[]eq}). 

\subsection{Israel junction condition}
We implicitly assume Einstein gravity in the bulk,
\begin{eqnarray}
{\cal G}_{AB}=\kappa_{(n)}^2T_{AB}\,,
\end{eqnarray}
where $\kappa_{(n)}^{2}$ is the $n$-dimensional
gravitational coupling constant.
Since we focus on the effective Einstein equation
on the brane, it is unnecessary to specify the matter in the bulk
and solve for the bulk geometry in the treatment below.
We simply assume that the bulk geometry is defined consistently.

The Israel junction condition~\cite{Israel:1966rt} is
\begin{eqnarray}
[K\gamma_{ab}-K_{ab}]=\kappa_{(n)}^{2}\bar{T}_{ab}\,,
\label{eq:Israel}
\end{eqnarray}
where $\bar{T}_{ab}$ is the energy-momentum tensor of the matter
on the brane. We decompose the energy-momentum tensor as
\begin{eqnarray}
\bar{T}_{ab}&&=-\lambda\gamma_{ab}+\tau_{ab}\,,
\label{eq:stress-energy decomposition}
\end{eqnarray}
where $\lambda$ is the brane tension.
Using this decomposition, we rewrite the junction condition as 
\begin{eqnarray}
[K_{ab}]=-\kappa_{(n)}^{2}
\left(\tau_{ab}+\frac{\lambda -\tau}{n-2}\gamma_{ab}\right)\,.
\label{eq:junction condition}
\end{eqnarray}
Using this junction condition,
the jump of the Gauss equation~(\ref{eq:Gauss eq.}) is written
\begin{eqnarray}
-\frac{1}{\kappa_{(n)}^{2}}[{\cal F}_{ab}]
=2\tau_{(a}^{c}\langle K_{b )c}\rangle
-\frac{(n-3)\lambda-\tau}{n-2}\langle K_{ab}\rangle
-\langle K\rangle\left(\tau_{ab}+\frac{\lambda-\tau}{n-2}\gamma_{ab}\right) ,
\label{eq:2ndGauss eq.}
\end{eqnarray}
where $\tau=\tau^a{}_a$. 
We note that this is the key equation that relates
the mean extrinsic curvature $\langle K_{ab}\rangle$ 
to the bulk curvature and the brane energy-momentum tensor.
As discussed below, a preliminary form of 
the effective Einstein equation on the brane,
which we call the pre-effective Einstein equation
on the brane, contains many terms involving
 $\langle K_{ab}\rangle$. In \S~\ref{sec:III},
we solve this equation for the mean extrinsic curvature
to obtain a meaningful effective Einstein equation on the brane.

In passing, we mention a necessary condition for the brane to be
embedded in the bulk in a consistent manner~\cite{Battye:2001pb}. Taking the
trace of Eq.~(\ref{eq:2ndGauss eq.}) or Eq.~(\ref{eq:Gauss eq.})
with the Israel junction condition~(\ref{eq:Israel}), we obtain
\begin{eqnarray}
[{\cal F}]=2\kappa_{(n)}^{2}\bar{T}^{ab}\langle K_{ab}\rangle\,.
\label{eq:F-f eq}
\end{eqnarray}
Comparing this with Eq.~(\ref{eq:F trace}), we obtain 
a part of the relation between the brane and the bulk, 
\begin{eqnarray}
2\kappa_{(n)}^{2}\bar{T}^{ab}\langle K_{ab}\rangle
 =[{\cal G}_{MN}]n^{M}n^{N}.
\label{eq:[TK]}
\end{eqnarray}
We mention that Israel derived this equation in the case
$[{\cal F}]=2[{\cal G}_{MN}]n^Mn^N=0$~\cite{Israel:1966rt}.

\subsection{Pre-effective Einstein equation on the brane}
\label{sec:effective Einstein eq.}
Using the results of the previous subsection,
we construct the effective Einstein equation on the brane
without any symmetry.
For convenience, we also decompose the extrinsic curvature into
its trace and traceless parts,
\begin{eqnarray}
&&K_{ab}=\frac{K}{n-1}\gamma_{ab}+\sigma_{ab}\,,
\label{eq:K decomposition}
\end{eqnarray}
where the traceless part, $\sigma_{ab}$, describes the shear
of the normal vector $n^M$, and the trace, $K$, describes its expansion.

The effective Einstein equation is derived from the mean
of the Gauss equation~(\ref{eq:<Gauss eq.>}). 
Inserting the junction condition~(\ref{eq:junction condition}) into it,
together with the decomposition formulas~(\ref{eq F_ab decomposition}) 
and (\ref{eq:K decomposition}) for ${\cal F}_{ab}$ and $K_{ab}$,
respectively, we obtain
\begin{eqnarray}
\bar{G}_{ab}=-\bar{\Lambda}\gamma_{ab}+\kappa_{(n-1)}^{2}\tau_{ab}
+\frac{\kappa_{(n-1)}^{2}}{\lambda}S_{ab}
+\langle \Omega_{ab}\rangle 
-\langle \sigma_{a}{}^{c}\rangle\langle  \sigma_{bc}\rangle\,,
\label{eq:Einstein eq.}
\end{eqnarray} 
where the effective $(n-1)$-dimensional gravitational constant
is defined by
\begin{eqnarray}
\kappa_{(n-1)}^{2}=\frac{n-3}{4(n-2)}\kappa_{(n)}^{4}\lambda\,,
\end{eqnarray}
and the other terms are given by
\begin{eqnarray}
&&\bar{\Lambda}=\frac{n-3}{2(n-1)}\langle {\cal F}\rangle 
+\frac{1}{2}\kappa_{(n-1)}^{2}\lambda
 +\frac{(n-2)(n-3)}{2(n-1)^{2}}\langle K \rangle^{2}
-\frac{1}{2}\langle \sigma^{cd} \rangle\langle \sigma_{cd} \rangle\,,
\label{eq:effective cosmological constant}
\\
&&S_{ab}=\frac{\tau}{n-3}\tau_{ab}
-\frac{\tau^{2}}{2(n-3)}\gamma_{ab}
-\frac{n-2}{n-3}\tau^{c}_{a}\tau_{bc}
+\frac{n-2}{2(n-3)}\tau^{cd}\tau_{cd}\gamma_{ab}\,,
\\
&&\langle \Omega_{ab}\rangle =\langle \omega_{ab}\rangle
 +\frac{n-3}{n-1}\langle K \rangle\langle \sigma_{ab} \rangle\,.
\end{eqnarray}
The first term $\bar{\Lambda}$ on the right-hand side of
Eq.~(\ref{eq:Einstein eq.}) represents the effective cosmological constant.
We note, however, that this quantity may not be constant in general,
as is clear from its expression in Eq.~(\ref{eq:effective cosmological constant}).
It reduces to a constant, for example,
in the $Z_2$ symmetric case, in which all the angled brackets except
for ${\cal F}_{ab}$ are zero.
The second and third terms are contributions from
the energy-momentum tensor on the brane and its quadratic term,
which are the same as in the $Z_2$ symmetric case~\cite{Shiromizu:1999wj}.
The fourth traceless term, $\langle\Omega_{ab}\rangle$, which is
traceless by definition, is an extension of the ${\cal E}_{ab}$ term
in the $Z_2$ symmetric case. Finally, the last term is a new term,
which has no analog in the $Z_2$ symmetric case.
As is clear from its definition, this term arises from 
the square of the mean extrinsic curvature
$\langle K_{ab}\rangle$, and it vanishes only 
if the traceless part of $\langle K_{ab}\rangle$ is zero.

The above effective Einstein equation,~(\ref{eq:Einstein eq.}), is 
completely general in the sense that no symmetry has been imposed.
However, it is useless, except in the $Z_2$ symmetric case,
because it depends strongly on the unknown mean extrinsic curvature
$\langle K_{ab}\rangle$:
It is contained in $\bar\Lambda$, $\langle\Omega_{ab}\rangle$ and
 $\langle\sigma_{a}{}^{c}\rangle\langle\sigma_{bc}\rangle$.
For this reason, we call it the pre-effective Einstein equation rather than
the effective Einstein equation.
In order to make it meaningful, it is necessary to express
 $\langle K_{ab}\rangle$ in terms of geometrical quantities
in the bulk (i.e., the bulk metric and curvature) and the brane 
energy-momentum tensor. This problem is addressed in the next
section.

\section{The mean extrinsic curvature $\langle K_{ab}\rangle$}
\label{sec:III}
\subsection{Junction condition for $\langle K_{ab}\rangle$}

The equation we solve is Eq.~(\ref{eq:2ndGauss eq.}),
which we present below for convenience:
\begin{eqnarray}
-\frac{1}{\kappa_{(n)}^{2}}[{\cal F}_{ab}]
=2\tau_{(a}^{c}\langle K_{b )c}\rangle
-\frac{(n-3)\lambda-\tau}{n-2}\langle K_{ab}\rangle
-\langle K\rangle\left(\tau_{ab}+\frac{\lambda-\tau}{n-2}\gamma_{ab}\right)\,.
\label{eq:new Gauss eq.}
\end{eqnarray}
The task is to solve this equation for $\langle K_{ab}\rangle$.
We call this the second Gauss equation.
Battye et al.~\cite{Battye:2001yn} solved this equation in two 
particular cases. One is the case in which
the elements of the brane energy-momentum tensor, $\tau_{ab}$, are sufficiently smaller
than the brane tension, $\lambda$, so that the low energy expansion
is valid. The other is the case in which the brane energy-momentum
tensor takes the perfect fluid form. Here we seek a
general solution without particular assumptions
concerning the brane energy-momentum tensor.

For convenience, we introduce the ``hatted" energy-momentum tensor,
$\hat\tau_{ab}$, which differs from the original one by a trace term :
\begin{eqnarray}
\hat{\tau}_{ab}\equiv
\tau_{ab}-\frac{(n-3)\lambda-\tau}{2(n-2)}\gamma_{ab}
=\bar{T}_{ab}-\frac{1}{2(n-2)}\bar{T}\gamma_{ab}\,.
\end{eqnarray}
Using this, the junction condition~(\ref{eq:junction condition})
becomes
\begin{eqnarray}
[K_{ab}]=\kappa_{(n)}^{2}
\left(-\hat{\tau}_{ab}+\frac{\hat{\tau}}{n-3}\gamma_{ab}\right)\,,
\label{eq:hat juction condition}
\end{eqnarray}
and the second Gauss equation,~(\ref{eq:new Gauss eq.}), 
becomes
\begin{eqnarray}
-[{\cal F}_{ab}]
=2\kappa_{(n)}^{2}\hat{\tau}_{(a}{}^{c}\langle K_{b)c}\rangle
 +\langle K\rangle [K_{ab}]\,.
\label{eq:general Gauss eq.}
\end{eqnarray}

Below we solve Eq.~(\ref{eq:general Gauss eq.}) to obtain the mean extrinsic curvature.
Our method consists of two parts. First, we obtain the trace of
the mean extrinsic curvature $\langle K\rangle$ by introducing
the inverse of the hatted tensor $\hat\tau_{ab}$.
Second, with $\langle K\rangle$ known, we rewrite the second
Gauss equation as a matrix equation for $\langle K_{ab}\rangle$.
This matrix equation can be solved by using the tetrad
(more precisely, the vielbein) decomposition of $\hat\tau_{ab}$. 

First, let us solve for the trace of the mean extrisic curvature.
We regard $\hat{\tau}_{ab}$ as an $(n-1)\times(n-1)$ symmetric matrix
and introduce its inverse matrix, $(\hat{\tau}^{-1})^{ab}$, assuming
$\det (\hat{\tau}_{ab})\neq0$. 
By definition, the inverse matrix $(\hat{\tau}^{-1})^{ab}$ satisfies
\begin{eqnarray}
(\hat{\tau}^{-1})^{ac}\hat{\tau}_{cb}=\delta^a{}_{b}\,.
\label{eq:tau^{-1} satisfy}
\end{eqnarray}
We multiply the second Gauss equation~(\ref{eq:general Gauss eq.}) 
by this inverse matrix $(\hat{\tau}^{-1})^{ab}$ and take the trace
to obtain
\begin{eqnarray}
\kappa_{(n)}^{2}\langle K\rangle
 =\frac{(n-3)(\hat{\tau}^{-1})^{ab}[{\cal F}_{ab}]}
{(n-3)^{2}-\hat{\tau}^{m}{}_{m}(\hat{\tau}^{-1})^{n}{}_{n}}\,,
\label{eq:general trace part}
\end{eqnarray}
where Eq.~(\ref{eq:hat juction condition}) has been used.
We note that the denominator is zero if
 $\hat{\tau}^{a}{}_{a}(\hat{\tau}^{-1})^{b}{}_{b}=(n-3)^{2}$, and thus
a special consideration is needed in this case.
We defer this to a future study.

Inserting the solution~(\ref{eq:general trace part}) into 
the second Gauss equation~(\ref{eq:general Gauss eq.}),
we obtain the matrix equation
\begin{eqnarray}
2\kappa_{(n)}^{2}\hat{\tau}_{(a}{}^{c}\langle K_{b)c}\rangle
 =-[\tilde{{\cal F}}_{ab}] \,,
\label{eq:master eq.}
\end{eqnarray}
where $[\tilde{\cal F}_{ab}]$ is defined by
\begin{eqnarray}
[\tilde{{\cal F}}_{ab}]\equiv
 [{\cal F}_{ab}]+\langle K\rangle [K_{ab}]
=[{\cal F}_{ab}]
+\frac{(n-3)(\hat{\tau}^{-1})^{cd}[{\cal F}_{cd}]}{(n-3)^{2}
-\hat{\tau}^{m}{}_{m}(\hat{\tau}^{-1})^{n}{}_{n}}
\left(-\hat{\tau}_{ab}+\frac{\hat{\tau}}{n-3}\gamma_{ab}\right)\,.\nonumber\\
\label{eq:F^tilde eq.}
\end{eqnarray}

To solve Eq.~(\ref{eq:master eq.}), we use the vielbein decomposition.
First, we introduce a set of orthonormal basis vectors
$e_{a}^{(i)}$ ($i=0,1,\cdots,n-2$),
where $e_{a}^{(0)}$ is timelike and the rest are spacelike.
They satisfy the orthonormality conditions
\begin{eqnarray}
&&e^{a}_{(i)}e_{a}{}_{(j)}=\eta_{(i)(j)}\,,
\label{eq:tetrad condition1}
\\
&&\sum_{i,j=0}^{n-2}\eta_{(i)(j)}e_{a}^{(i)}e_{b}^{(j)}
=\sum_{j=0}^{n-2}e_{a}^{(j)}e_{b(j)}=\gamma_{ab}\,,
\label{eq:tetrad condition2}
\end{eqnarray}
where $\eta_{(i)(j)}$ is the Minkowski metric.
We note that here we do not adopt the standard Einstein convention 
for the summation over the frame indices $(i)$; instead,
we explicitly insert the symbol $\sum$ when 
the summation is to be taken.

Because the matrix $\hat{\tau}_{ab}$ is symmetric,
it can be diagonalized by an orthogonal matrix $P_{ab}$ : 
\begin{eqnarray}
(P^{-1})_{a}{}^{c}\hat{\tau}_{cd}P^{d}{}_{b}
=\sum_{i}\hat{\tau}_{(i)}e_{a}^{(i)}e_{b(i)}\,.
\end{eqnarray}
We note that $P_{ab}$ 
represents a local Lorentz transformation of the frame.
Thus the frame determined by $\bar{e}_{a}^{(i)}\equiv P_{a}{}^{c}e_{c}^{(i)}$ 
corresponds to a local Lorentz frame in which the hatted tensor $\hat\tau_{ab}$
(and hence the original energy-momentum tensor $\tau_{ab}$) is diagonalized :
\begin{eqnarray}
\hat\tau_{ab}
=\sum_{i}\hat{\tau}_{(i)}\bar{e}_{a}^{(i)}\bar{e}_{b(i)}\,.
\label{eq:diagonal tau}
\end{eqnarray}
Inserting this into the matrix equation~(\ref{eq:master eq.})
and contract in it with $\bar{e}^{a}_{(i)}\bar{e}^{b}_{(j)}$, 
we obtain
\begin{eqnarray}
\kappa_{(n)}^{2}\left(\hat{\tau}_{(i)}+\hat{\tau}_{(j)}\right)
\bar{e}_{(i)}^{a}\bar{e}_{(j)}^{b}\langle K_{ab}\rangle
 =-\bar{e}_{(i)}^{a}\bar{e}_{(j)}^{b}[\tilde{{\cal F}}_{ab}]\,,
\end{eqnarray}
or equivalently,
\begin{eqnarray}
\kappa_{(n)}^{2}\left(\hat{\tau}_{(i)}+\hat{\tau}_{(j)}\right)
\langle K_{(i)(j)}\rangle =-[\tilde{{\cal F}}_{(i)(j)}]\,,
\end{eqnarray}
where $K_{(i)(j)}$ and $F_{(i)(j)}$ denote the vielbein components.
We recall that this equation holds for each pair of $(i)$ and $(j)$.
Then, we divide both sides by $\hat{\tau}_{(i)}+\hat{\tau}_{(j)}$,
assuming this is non-zero, and obtain
\begin{eqnarray}
\kappa_{(n)}^{2}\langle K_{ab}\rangle
 = -\sum_{i,j}\frac{1}{\hat{\tau}_{(i)}+\hat{\tau}_{(j)}}
\bar{e}_{a}^{(i)}\bar{e}_{b}^{(j)}[\tilde{{\cal F}}_{(i)(j)}]\,.
\end{eqnarray}
We also note that this result is valid only if we have 
$\hat{\tau}_{(i)}+\hat{\tau}_{(j)}\neq0$ for all possible
pairs of $(i)$ and $(j)$. We need a special treatment
in the case that any of the demonimators are zero.
Again, we defer treatment this special case to a future study.

The diagonal term can be written
\begin{eqnarray}
\sum_{i=j}\frac{1}{\hat{\tau}_{(i)}+\hat{\tau}_{(j)}}\bar{e}_{a}^{(i)}
\bar{e}_{b}^{(j)}[\tilde{{\cal F}}_{(i)(j)}]
=\frac{1}{2}(\hat{\tau}^{-1})_{a}{}^{c}[\tilde{{\cal F}}_{cb}]\,.
\label{eq:diag}
\end{eqnarray}
Using this, we arrive at the final result,
\begin{eqnarray}
\kappa_{(n)}^{2}\langle K_{ab}\rangle
 &=&-\frac{1}{2}(\hat{\tau}^{-1})_{a}{}^{c}[{\cal F}_{bc}]
+\frac{(n-3)(\hat{\tau}^{-1})^{de}[{\cal F}_{de}]}
{2( (n-3)^{2}-\hat{\tau}^{m}{}_{m} (\hat{\tau}^{-1})^{n}{}_{n})}
\left(\gamma_{ab}-\frac{\hat{\tau}^{c}{}_{c}}{n-3}(\hat{\tau}^{-1})_{ab}\right)
\nonumber\\
&&\qquad
 -\sum_{i\neq j}\frac{1}{\hat{\tau}_{(i)}+\hat{\tau}_{(j)}}
\bar{e}_{a}^{(i)}\bar{e}_{b}^{(j)}[\omega_{(i)(j)}]\,.
\label{eq:general solution}
\end{eqnarray}
We refer to this as the junction condition for the mean
of the extrinsic curvature, which is a counterpart to the conventional
junction condition for the jump of the extrinsic curvature,
Eq.~(\ref{eq:junction condition}).
If convenient, the above result can be decomposed into 
its trace and traceless parts,
\begin{eqnarray}
\kappa_{(n)}^{2}\langle K\rangle
 &=&\frac{(n-3)(\hat{\tau}^{-1})^{ab}[{\cal F}_{ab}]}{(n-3)^{2}
-\hat{\tau}^{m}{}_{m}(\hat{\tau}^{-1})^{n}{}_{n}}\,,
\\
\kappa_{(n)}^{2}\langle\sigma_{ab}\rangle
 &=&-\frac{1}{2}(\hat{\tau}^{-1})_{a}{}^{c}[{\cal F}_{bc}]\nonumber\\
&&+\frac{(\hat{\tau}^{-1})^{cd}[{\cal F}_{cd}]}{2\{ (n-3)^{2}
-\hat{\tau}^{m}{}_{m}(\hat{\tau}^{-1})^{n}{}_{n}\}}
\left(\frac{(n-3)^{2}}{n-1}\gamma_{ab}
-\hat{\tau}^{e}{}_{e}(\hat{\tau}^{-1})_{ab}\right)
 \nonumber\\
&&-\sum_{i\neq j}\frac{1}{\hat{\tau}_{(i)}
+\hat{\tau}_{(j)}}\bar{e}_{a}^{(i)}\bar{e}_{b}^{(j)}[\omega_{(i)(j)}]\,.
\end{eqnarray}

To conclude this subsection, we have obtained 
the general solution for the mean extrinsic curvature 
expressed in terms of the bulk curvature and the brane energy-momentum.
The result constitutes the junction condition for the mean of the extrinsic
curvature across the brane, and it is analogous to the well-known junction condition
for the jump of the extrinsic curvature. As we have noted, this result
is valid in all but a few cases, in which the equation
to be solved is singular. A careful analysis of these singular cases
will be given in the future.

\subsection{Effect of $\langle K_{ab}\rangle$ on the brane}
In the previous subsection, we obtained the junction
condition for the mean extrinsic curvature.
In this subsection, we discuss its effect on the brane
from several points of view.

\subsubsection{Effectively anisotropic fluid}
In the frame of the basis vectors $\bar{e}_{a}^{(i)}(=P_{a}{}^{b}e_{b}^{(i)})$,
we have the diagonalized energy-momentum tensor
\begin{eqnarray}
\tau_{ab}=\sum \tau_{(i)}\bar{e}_{a}^{(i)}\bar{e}_{b(i)}\,,
\end{eqnarray}
where $\tau_{(i)}=\tau_{ab}\bar{e}_{(i)}^{a}\bar{e}^{b(i)}$,
and the quadratic term $S_{ab}$ is given by
\begin{eqnarray}
S_{ab}\bar{e}_{(i)}^{a}\bar{e}_{(j)}^{b}
&=&\left(\frac{1}{n-3}\tau\tau_{(i)}-\frac{\tau^{2}}{2(n-3)}
-\frac{n-3}{n-2}\tau_{(i)}^{2}+\frac{n-2}{2(n-3)}
\left(\sum_{k}\tau_{(k)}^{2}\right)\right)
\eta_{(i)(j)} 
\nonumber\\
&\equiv& S_{(i)}\eta_{(i)(j)}\,.
\end{eqnarray}
With these, we can rewrite the effective Einstein equation 
in terms of the vielbein components.

The diagonal components of the Einstein equation are
\begin{eqnarray}
\bar{G}_{(j)(j)}
=-\bar{\Lambda}+\kappa_{(n-1)}^{2}\tau_{(j)}
+\kappa_{(n-1)}^{2}S_{(j)}+\langle \Omega_{(j)(j)}\rangle
 +\sum_{k}\langle\sigma_{(j)}{}^{(k)}\rangle\langle\sigma_{(j)(k)}\rangle\,,
\end{eqnarray}
for each $(j)$, and the off-diagonal components are
\begin{eqnarray}
\bar{G}_{(i)(j)}
=\langle \Omega_{(i)(j)}\rangle
 +\sum_{k}\langle\sigma_{(i)}{}^{(k)}\rangle\langle\sigma_{(k)(j)}\rangle\,,
\label{eq:non-diagonal Einstein eq.}
\end{eqnarray}
for $(i)\neq(j)$.
The latter expression shows that
the source terms for the off-diagonal components
originate from $\langle\Omega_{(i)(j)}\rangle$ and 
$\langle\sigma_{(i)(j)}\rangle$. 
This should be compared to the $Z_2$ symmetric case, in which
the off-diagonal components are due only to 
${\cal E}_{ab}$~\cite{Maartens:2003tw}.
In the present case, in addition to $\Omega_{ab}$,
which is a generalization of ${\cal E}_{ab}$,
there is a contribution to off-diagonal components
from the new term,
$\langle\sigma_{a}{}^{c}\rangle\langle\sigma_{cb}\rangle$.

\subsubsection{Low energy limit}
We conjecture that the low energy regime, where $|\tau_{ab}|\ll\lambda$,
Einstein gravity is recovered on the brane.
For this reason, we believe that the contributions of the $\omega_{ab}$ and $S_{ab}$ terms
become negligibly small. To examine this,
let us consider the solution for the mean extrinsic curvature
up to ${O}(\tau_{ab}^{2},\omega_{ab})$.
In this case, the inverse of the hatted energy-momentum tensor
is given by
\begin{eqnarray}
(\hat\tau^{-1})^{ab}
=\frac{2(n-2)}{n-3}\lambda^{-1}
\left(\gamma^{ab}
+\frac{2(n-2)}{n-3}\lambda^{-1}
\left(\tau^{ab}-\frac{\tau}{2(n-2)}\gamma^{ab}\right)
+\cdots\right)\,.\nonumber\\
\end{eqnarray}
Then $\langle K_{ab}\rangle$ can be readily obtained as
\begin{eqnarray}
\kappa_{(n)}^{2}\langle K_{ab}\rangle
 =\frac{[{\cal F}]}{2(n-1)\lambda}
\left\{\gamma_{ab}+\frac{1}{\lambda}
\left(\tau\gamma_{ab}-(n-2)\tau_{ab}\right)\right\}
 +{\cal O}(\tau_{ab}^{2},\omega_{ab})\,.
\end{eqnarray}
Using this, the effective Einstein equation becomes
\begin{eqnarray}
\bar{G}_{ab}
=-\bar{\Lambda}^{LE}\gamma_{ab}+\kappa_{(n-1)}^{LE}\tau_{ab}
+{O}(\tau_{ab}^{2},\omega_{ab}) ,
\end{eqnarray}
where 
\begin{eqnarray}
&&\kappa_{(n-1)}^{LE}=\frac{n-3}{4(n-2)}\kappa_{(n)}^{4}\lambda
 -\frac{(n-2)(n-3)[{\cal F}]^{2}}{4(n-1)^{2}\kappa_{(n)}^{4}\lambda^{3}}\,,
\\
&&\bar{\Lambda}^{LE}
=\frac{n-3}{2(n-1)}\langle {\cal F}\rangle 
+\frac{n-3}{8(n-2)}(\kappa_{(n)}^{2}\lambda )^{2}
+\frac{(n-2)(n-3)[{\cal F}]^{{2}}}{8\kappa_{(n)}^{4}(n-1)^{2}\lambda^{2}}\,.
\end{eqnarray}
Thus, Einstein gravity is recovered. 
However, in contrast to our naive expectation, the contribution of the mean
extrinsic curvature gives rise to new correction terms from the bulk,
both to the gravitational constant and to the cosmological constant,
which are not necessarily constant.

\subsubsection{Perfect fluid case}
We next consider the case of a perfect fluid, in which we have
\begin{eqnarray}
\tau_{ab}=\rho u_{a}u_{b}+p(\gamma_{ab}+u_{a}u_{b})\,.
\end{eqnarray} 
In this case, it is convenient to decompose $\omega_{ab}$ as
\begin{eqnarray}
\omega_{ab}=\kappa_{(n)}^{2}\left\{ \rho^{\omega}u_{a}u_{b}
+\frac{1}{n-2}\rho^{\omega}(\gamma_{ab}+u_{a}u_{b})
+2q^{\omega}_{(a}u_{b)}+\pi^\omega_{ab}\right\}\,.
\end{eqnarray}
Then we obtain
\begin{eqnarray}
\kappa_{(n)}^{2}\langle K_{ab}\rangle 
&=&-\frac{[{\cal F}]}{2(n-1)(\rho +\lambda )^{2}}
\{ (\rho +p)u_{a}u_{b}+(\rho +\lambda )\gamma_{ab}\}
\nonumber\\
&&+\frac{n-2}{(n-3)(\rho +\lambda )}
\left\{ -\frac{p-\lambda }{\rho +\lambda}[\rho^{\omega}]u_{a}u_{b}
+\frac{1}{n-2}[\rho^{\omega}](\gamma_{ab}+u_{a}u_{b})
+2[q^{\omega}_{(a}]u_{b)}\right\}
\nonumber\\
&&-\frac{n-2}{\rho +(n-2)p-(n-3)\lambda}[\pi^\omega_{ab}] \,.
\end{eqnarray}
This is consistent with the result of Battye et al.~\cite{Battye:2001yn}.

\section{Conclusion}
\label{sec:IV}
In this paper, we considered a general codimension-1 brane
in an arbitrary number of dimensions without $Z_2$ symmetry
and derived the junction conditions in a complete form.
More specifically, we obtained expressions for both the
jump and the mean of the extrinsic curvature in terms of
the bulk curvature tensor and the brane energy-momentum tensor.
With this result, we derived the effective Einstein equation
on the brane in its most general form,
which is a generalization 
of the Shiromizu-Maeda-Sasaki equation~\cite{Shiromizu:1999wj}
to the case in which $Z_{2}$ symmetry does not exist.

The derived effective Einstein equation has a new term 
arising from the mean extrinsic curvature, 
and this new term leads to the appearance effectively 
anisotropic matter on the brane.
This gives rise to the possibility of
anisotropic brane worlds with 
an anisotropic bulk geometry, even in the case of
an isotropic brane energy-momentum tensor.

Thus, our result is a basic equation for the hybrid brane world
scenario, in which some spatial dimensions on the brane are
Kaluza-Klein compactified~\cite{Chatillon:2006vw}. 
Also, it provides a basis for higher codimension brane worlds
in which a higher codimension brane is regularized by
a codimension-1 brane with extra dimensions on the 
brane compactified to an infinitesimally small size.

There is, however, one subtlety in our derivation.
We had to assume the energy-momentum tensor on the brane to satisfy
several conditions. When these conditions are not satisfied, the
equation for the mean extrinsic curvature is singular. We plan
to investigate this singular case in a future publication.

\section*{Acknowledgements}
We would like to thank M. Minamitsuji for valuable and interesting discussions. 
We would also like to thank Y. Sendouda, and K. Izumi for useful comments. 
This work was supported in part by 
the Monbu-Kagakusho 21st century COE Program 
``Center for Diversity and Universality in Physics",
and by JSPS Grants-in-Aid for Scientific Research 
[(B) No.~17340075, and (A) No.~18204024.]

\appendix
\section{Energy-Momentum Conservation}
Here, for reference, we evaluate the energy-momentum
conservation law on the brane. From the Codazzi
equation~(\ref{eq:Codazzi eq.}), we have
\begin{eqnarray}
\bar{\nabla}_{b}K-\bar{\nabla}_{a}K^{a}{}_{b}
={\cal R}_{MN}n^{M}\gamma^{N}{}_{b}={\cal G}_{MN}n^{M}\gamma^{N}{}_{b}\,.
\end{eqnarray}
Appying the junction condition~(\ref{eq:junction condition})
to the above, we obtain
\begin{eqnarray}
&&\bar{\nabla}_{a}\tau^{a}{}_{b}
=-[{\cal R}_{MN}]n^{M}\gamma^{N}{}_{b}\,,
\label{eq:energy conservation1}
\\
&&\bar{\nabla}_{b}\langle K\rangle
-\bar{\nabla}_{a}\langle K^{a}{}_{b}\rangle
=\frac{n-2}{n-1}\bar{\nabla}_{b}\langle K\rangle
 -\bar{\nabla}_{a}\langle \sigma^{a}{}_{b}\rangle
 =\langle {\cal R}_{MN}\rangle n^{M}\gamma^{N}{}_{b}\,,
\label{eq:energy conservation2}
\end{eqnarray}
where $\sigma$ is the traceless part of $K_{ab}$.
The first equation reveals
the violation of the energy-momentum conservation 
in the case that there is a jump in the bulk Ricci tensor.
The second equation
describes the exchange of the effective energy-momentum
due to the mean extrinsic curvature and the bulk 
energy-momentum.


\begin{thebibliography}{99}
%%%%%%%%%%%%%%%%%%%%%%%%%%%%%%%%%%%%%%%%%%%%%%%%%%%%%%%%%%%%%
% Some macros are available for the bibliography:
%  o for general use
%    \JL : general journals                 \andvol : Vol (Year) Page
%  o for individual journal 
%    \AJ   : Astrophys. J.           \NC         : Nuovo Cim.
%    \ANN  : Ann. of Phys.           \NPA, \NPB  : Nucl. Phys. [A,B]
%    \CMP  : Commun. Math. Phys.     \PLA, \PLB  : Phys. Lett. [A,B]
%    \IJMP : Int. J. Mod. Phys.      \PRA - \PRE : Phys. Rev. [A-E]     
%    \JHEP : J. High Energy Phys.    \PRL        : Phys. Rev. Lett.
%    \JMP  : J. Math. Phys.          \PRP        : Phys. Rep.
%    \JP   : J. of Phys.             \PTP        : Prog. Theor. Phys.     
%    \JPSJ : J. Phys. Soc. Jpn.      \PTPS       : Prog. Theor. Phys. Suppl.
% Usage:
%  \PRD{45,1990,345}          ==> Phys.~Rev.\ \textbf{D45} (1990), 345
%  \JL{Nature,418,2002,123}   ==> Nature \textbf{418} (2002), 123
%  \andvol{B123,1995,1020}    ==> \textbf{B123} (1995), 1020
%%%%%%%%%%%%%%%%%%%%%%%%%%%%%%%%%%%%%%%%%%%%%%%%%%%%%%%%%%%%%
  
%\cite{Randall:1999vf}
\bibitem{Randall:1999vf}
  L.~Randall and R.~Sundrum,
  %``An alternative to compactification,''
  Phys.\ Rev.\ Lett.\  {\bf 83} (1999), 4690
 ; hep-th/9906064.
  %%CITATION = PRLTA,83,4690;%%
%\cite{Randall:1999ee}
\bibitem{Randall:1999ee}
  L.~Randall and R.~Sundrum,
  %``A large mass hierarchy from a small extra dimension,''
  Phys.\ Rev.\ Lett.\  {\bf 83} (1999), 3370
 ; hep-ph/9905221.
  %%CITATION = PRLTA,83,3370;%%
%\cite{Horava:1996ma}
\bibitem{Horava:1996ma}
  P.~Horava and E.~Witten,
  %``Eleven-Dimensional Supergravity on a Manifold with Boundary,''
  Nucl.\ Phys.\  B {\bf 475} (1996), 94
 ; hep-th/9603142.
  %%CITATION = NUPHA,B475,94;%%
  %\cite{Cline:2003ak}
\bibitem{Cline:2003ak}
  J.~M.~Cline, J.~Descheneau, M.~Giovannini and J.~Vinet,
  %``Cosmology of codimension-two braneworlds,''
  J. High Energy Phys. {\bf 06} (2003), 048
 ; hep-th/0304147.
  %%CITATION = JHEPA,0306,048;%%
%\cite{Chatillon:2006vw}
\bibitem{Chatillon:2006vw}
  N.~Chatillon, C.~Macesanu and M.~Trodden,
  %``Brane cosmology in an arbitrary number of dimensions,''
  Phys.\ Rev.\  D {\bf 74} (2006), 124004
 ; gr-qc/0609093.
  %%CITATION = PHRVA,D74,124004;%%
%\cite{Peloso:2006cq}
\bibitem{Peloso:2006cq}
  M.~Peloso, L.~Sorbo and G.~Tasinato,
  %``Standard 4d gravity on a brane in six dimensional flux compactifications,''
  Phys.\ Rev.\  D {\bf 73} (2006), 104025
 ; hep-th/0603026.
  %%CITATION = PHRVA,D73,104025;%%
%\cite{Kobayashi:2007kv}
\bibitem{Kobayashi:2007kv}
  T.~Kobayashi and M.~Minamitsuji,
  %``Gravity on an extended brane in six-dimensional warped flux
  %compactifications,''
  Phys.\ Rev.\  D {\bf 75} (2007), 104013
 ; hep-th/0703029.
  %%CITATION = PHRVA,D75,104013;%%
%\cite{Papantonopoulos:2006dv}
\bibitem{Papantonopoulos:2006dv}
  E.~Papantonopoulos, A.~Papazoglou and V.~Zamarias,
  %``Regularization of conical singularities in warped six-dimensional
  %compactifications,''
  J. High Energy Phys. {\bf 03} (2007), 002
 ; hep-th/0611311.
  %%CITATION = JHEPA,0703,002;%%
%\cite{Himmetoglu:2006nw}
\bibitem{Himmetoglu:2006nw}
  B.~Himmetoglu and M.~Peloso,
  %``Isolated Minkowski vacua, and stability analysis for an extended brane in
  %the rugby ball,''
  Nucl.\ Phys.\  B {\bf 773} (2007), 84
 ; hep-th/0612140.
  %%CITATION = NUPHA,B773,84;%%
%\cite{Kanno:2007wj}
\bibitem{Kanno:2007wj}
  S.~Kanno, D.~Langlois, M.~Sasaki and J.~Soda,
  %``Kaluza-Klein braneworld cosmology with static internal dimensions,''
  arXiv:0707.4510 [hep-th].
  %%CITATION = ARXIV:0707.4510;%%
%\cite{Shiromizu:1999wj}
\bibitem{Shiromizu:1999wj}
  T.~Shiromizu, K.~i.~Maeda and M.~Sasaki,
  %``The Einstein equations on the 3-brane world,''
  Phys.\ Rev.\  D {\bf 62} (2000), 024012
 ; gr-qc/9910076.
  %%CITATION = PHRVA,D62,024012;%%
%\cite{Battye:2001yn}
\bibitem{Battye:2001yn}
  R.~A.~Battye, B.~Carter, A.~Mennim and J.~P.~Uzan,
  %``Einstein equations for an asymmetric brane-world,''
  Phys.\ Rev.\  D {\bf 64} (2001), 124007
 ; hep-th/0105091.
  %%CITATION = PHRVA,D64,124007;%%
%\cite{Gergely:2003pn}
\bibitem{Gergely:2003pn}
  L.~A.~Gergely,
  %``Generalized Friedmann branes,''
  Phys.\ Rev.\  D {\bf 68} (2003), 124011
 ; gr-qc/0308072.
  %%CITATION = PHRVA,D68,124011;%%
%\cite{Maia:2004fq}
\bibitem{Maia:2004fq}
  M.~D.~Maia, E.~M.~Monte, J.~M.~F.~Maia and J.~S.~Alcaniz,
  %``On the geometry of dark energy,''
  Class.\ Quant.\ Grav.\  {\bf 22} (2005), 1623
 ; astro-ph/0403072.
  %%CITATION = CQGRD,22,1623;%%
%\cite{Ida:1999ui}
\bibitem{Ida:1999ui}
  D.~Ida,
  %``Brane-world cosmology,''
  J. High Energy Phys. {\bf 09} (2000), 014
 ; gr-qc/9912002.
  %%CITATION = JHEPA,0009,014;%%
%\cite{Kanno:2005vq}
\bibitem{Kanno:2005vq}
  S.~Kanno, J.~Soda and D.~Wands,
  %``Braneworld flux inflation,''
  J. Cosmol. Astropart. Phys. {\bf 08} (2005), 002
  [arXiv:hep-th/0506167].
  %%CITATION = JCAPA,0508,002;%%
%\cite{Israel:1966rt}
\bibitem{Israel:1966rt}
  W.~Israel,
  %``Singular hypersurfaces and thin shells in general relativity,''
  Nuovo Cim.\  B {\bf 44S10} (1966), 1
  [Errata, {\bf 48} (1967\ NUCIA,B44,1.1966) 463].
  %%CITATION = NUCIA,B44,1;%%
%\cite{Battye:2001pb}
\bibitem{Battye:2001pb}
  R.~A.~Battye and B.~Carter,
  %``Generic junction conditions in brane-world scenarios,''
  Phys.\ Lett.\  B {\bf 509} (2001), 331
 ; hep-th/0101061.
  %%CITATION = PHLTA,B509,331;%%
%\cite{Maartens:2003tw}
\bibitem{Maartens:2003tw}
  R.~Maartens,
  %``Brane-world gravity,''
  Living Rev.\ Relativity\  {\bf 7} (2004), 7
 ; gr-qc/0312059.
  %%CITATION = 00222,7,7;%%
\end{thebibliography}
\end{document}